\documentclass[preprint]{aastex}

\usepackage{epsfig}

\newcommand{\be}{\begin{equation}}
\newcommand{\ee}{\end{equation}}
\newcommand{\ba}{\begin{eqnarray}}
\newcommand{\ea}{\end{eqnarray}}

\newcommand{\bft}{\mbox{\boldmath $\theta$}}

\begin{document}

\pagestyle{plain}

\title{Diagnosing space telescope misalignment and jitter using stellar images}
\author{Zhaoming Ma{\footnote{Email: mazh@sas.upenn.edu}}, Gary Bernstein}
\affil{Department of Physics \& Astronomy,\\
       University of Pennsylvania, Philadelphia, PA 19104}

\author{Alan Weinstein}
\affil{Department of Physics, \\
       California Institute of Technology, Pasadena, CA 91125}
\and
\author{Michael Sholl}
\affil{Space Science Laboratory, \\
       University of California at Berkeley, Berkeley, CA 94720}

\begin{abstract}

Accurate knowledge of the telescope's point spread function (PSF) is
essential for the weak gravitational lensing measurements that hold
great promise for cosmological constraints.  For space telescopes, the
PSF may vary with time due to thermal drifts in the telescope
structure, and/or due to jitter in the spacecraft pointing
(ground-based telescopes have additional sources of variation). 
We describe and
simulate a procedure for using the images of the stars in each exposure to
determine the misalignment and jitter parameters, and reconstruct the
PSF at any point in that exposure's field of view.
The simulation uses the design of the
SNAP{\footnote{http://snap.lbl.gov}} telescope.
Stellar-image data in a typical exposure determines secondary-mirror
positions as precisely as $20\,{\rm nm}$. 
The PSF ellipticities and size, which are the quantities of interest for
weak lensing are determined to $4.0 \times 10^{-4}$ and  $2.2 \times 10^{-4}$
accuracies respectively in each exposure,
sufficient to meet weak-lensing requirements.  We show that, for the case of a
space telescope, the PSF estimation errors scale inversely with the
square root of the total number of photons collected from all the
usable stars in the exposure.

\end{abstract}
\keywords{cosmology -- gravitational lensing, large-scale structure of the
          universe}

\section{Introduction}
\label{sec:intro}

    The accelerated expansion of the universe is one of the most puzzling
astrophysical discovery of the century.
The proposed explanations are dark energy, modified gravity
and feedback from density fluctuations. To explore the mystery, a few
large astronomical surveys are underway
(DES{\footnote{http://www.darkenergysurvey.org}},
 ACT{\footnote{http://www.physics.princeton.edu/act}},
 SPT{\footnote{http://pole.uchicago.edu}})
or in planning stages (SNAP,
 DESTINY{\footnote{http://destiny.asu.edu}},
 LSST{\footnote{http://www.lsst.org}},
 EUCLID{\footnote{The merger of DUNE (http://www.dune-mission.net) and
                  SPACE \citep{SPACE}}}).
The sensitivity of these surveys to the expansion
of the universe comes from both cosmological distances and growth of density
perturbations as a function of the cosmic time or redshift. The probes 
utilized are Type-Ia supernova, weak gravitational lensing (WL), baryon
acoustic oscillations, galaxy cluster counting (selected by optical or using
Sunyave-Zeldovic effect), and the integrated Sachs-Wolfe effect (ISW).
Among all these probes, weak lensing is potentially the most rewarding
one if systematics are well under control. The dominating systematics
for weak lensing measurements include galaxy shape measurement errors
\citep{Huterer06,STEP1,STEP2, Stabenau07, Amara07},
photometric redshift errors \citep{Huterer06, Maetal06, MaBernstein08},
uncertainties of the matter power spectrum \citep{Huterer05, Bernstein08},
galaxy intrinsic alignment \citep{King02, King03, Heymans03, King05,
       Mandelbaum06,Heymans06,Hirata07, Bridle07, Lee08, Joachimi08},
and higher order effects such as reduced shear \citep{White05, Dodelson06,
       Shapiro08}, 
Born approximation \citep{Cooray02, Shapiro06},
and source clustering \citep{Bernardeau98, Hamana01, Schneider02}.

This paper is concerned with reducing the systematic errors in galaxy
shape measurements.  For future weak lensing surveys, the tolerable
RMS multiplicative calibration error on WL shear is about $10^{-3}$
\citep{Huterer06,Amara07}.
Additive errors in galaxy shear should also be held to $<10^{-3.5}$.
Mis-estimation of the PSF will propagate into systematic errors in the
shear.  The size of the PSF must therefore be determined to
better than 1 part in $10^3$ to avoid unacceptable multiplicative
shear error; likewise the PSF ellipticity must be known to $10^{-3}$ or
better to avoid unacceptable additive shear systematic
\citep{PaulinHenriksson}.

Unfortunately the PSF of real telescopes changes with time, as well as
with color and field position.
Effects that could change the PSF include thermal expansion of mirrors
and support structures; pointing jitter due to structural vibrations
and tracking errors; for ground-based telescopes there are
additionally gravity loading, atmospheric distortions, and wind
loading. 
% The engineers will try their best to reduce the above effects, but
% there are limits to what the engineers can do given finite resources.
Careful engineering of the telescope, mount, and housing can minimize
these effects, but there are limits to what can be achieved with finite
resources.
Even for a space telescope it is prohibitively expensive to guarantee
PSF stability to $<1$ part in $10^3$.  
On the other hand there will be stars in each image to diagnose the
PSF behavior during each exposure. Recall that the WL analysis
requires part-per-thousand {\em knowledge} of the PSF at each location
and each 
exposure, not necessarily that the PSF be {\em constant} to this level.
In this work, we study how well the
PSF can be constrained using stellar images, using the proposed
space-based SNAP telescope as a case study.

    The outline of the paper is as follows. We present an overview of the
task we are trying to accomplish in \S \ref{sec:task}. In \S \ref{sec:PSF},
we provide details of the modeling of the SNAP PSF. We describe the fitting
procedure for the misalignment and jitter parameters in \S \ref{sec:procedure}.
The results of the fit are presented in \S \ref{sec:result} and we conclude in
\S \ref{sec:conclude}.

%\section{Description of the Task}
\section{Stellar ``Morphometry"}
\label{sec:task}

   Weak lensing measurements aim to extract a map of the cosmic shear from
the coherent distortions in the shapes of many distant galaxies
\citep{Kaiser98, Bartelmann99}. Observed galaxy shapes are distorted by the
telescope PSF. To lowest order, this can be corrected if the PSF across
the telescope field of view is known. The PSF can be inferred from the
observed shapes of foreground stars of suitable magnitude. Because the PSF
will drift over time, it is desirable to measure the PSF across the field
of view for every exposure, using the stars that are interspersed throughout
the image along with the distant galaxies of interest. We refer to this
procedure as "stellar morphometry".

\subsection{Quantities of Interest}
  The requirements for a weak-lensing survey can be most simply stated
as limits on the tolerable error in the second moments of the PSF.  We
will measure the Gaussian-weighted moments defined as the zeroth moment
\be
    M_0 = \int dx\,dy\,PSF(x,y)W(x,y),
\label{eqn:m0}
\ee
the first moments
\be
   \bar{x} = {1 \over {M_0}} \int dx\,dy\,x\,PSF(x,y)W(x,y) \,,
\ee
\be
   \bar{y} = {1 \over {M_0}} \int dx\,dy\,y\,PSF(x,y)W(x,y) \,,
\ee
and the second moments
\be
   P_{xx} = {1 \over {M_0}} \int dx\,dy\,(x-\bar{x})^2 PSF(x,y)W(x,y) \,,
\ee
\be
   P_{yy} = {1 \over {M_0}} \int dx\,dy\,(y-\bar{y})^2 PSF(x,y)W(x,y) \,,
\ee
\be
   P_{xy} = {1 \over {M_0}} \int dx\,dy\,(x-\bar{x})(y-\bar{y}) PSF(x,y)W(x,y) \,.
\label{eqn:pxy}
\ee
These are calculated under a weight function which we take to be
\be
  W(x,y)={\rm exp} \left \{-{{(x-\bar{x})^2 + (y-\bar{y})^2} \over {2 \sigma_{mom}^2}} \right \} \,.
\ee
The width of the weighting Gaussian is chosen to be two
times the Airy radius,
\be
  \sigma_{mom} = 2 \times 1.22 {{\lambda f} \over {D}} \,,
\ee
where $\lambda$ is the wavelength of the incident light, $f$ is the focal
length and $D$ is the telescope aperture.
    In analogy, we can write down the third moments of the PSF,
$P_{xxx}$, $P_{xxy}$, $P_{xyy}$, and $P_{yyy}$.
%We leave details to the reader.

We also compute quantities derived from the second moments.
The ellipticities and stellar size are
\be
   e_1={{P_{xx} - P_{yy}} \over {P_{xx} + P_{yy}}} \,; \,
   e_2={{2 P_{xy}} \over {P_{xx} + P_{yy}}} \,; \,
   \sigma_\star^2={{P_{xx} + P_{yy}} \over {2}} \,.
\label{eqn:es}
\ee
These quantities appear in many approaches to weak-lensing shear
measurement [\cite{STEP1} and references therein]. The true PSF of an
exposure will depend on the field position of the star $(x_\star,y_\star)$
yielding $e_1(x_\star,y_\star)$, $e_2(x_\star,y_\star)$,
$\sigma^2_\star(x_\star,y_\star)$. Our goal is to produce an accurate model
estimate $\hat e_1(x_\star,y_\star),$ etc.  We will evaluate our success by
calculating the RMS residual errors in the ellipticity models,
\be
\left(e_1^{\rm RMS}\right)^2  \equiv 
\left\langle \left[e_1(x_\star,y_\star)-\hat e_1(x_\star,y_\star)\right]^2\right\rangle,
\label{eqn:e1RMS}
\ee
\be
\left(e_2^{\rm RMS}\right)^2  \equiv 
\left\langle \left[e_2(x_\star,y_\star)-\hat e_2(x_\star,y_\star)\right]^2\right\rangle,
\label{eqn:e2RMS}
\ee
and the fractional residual error in the PSF size,
\begin{equation}
\left(\sigma_\star^{\rm RMS}\right)^2  \equiv 
{ \left\langle \left[\sigma_\star(x_\star,y_\star)-\hat
      \sigma_\star(x_\star,y_\star)\right]^2\right\rangle \over \langle
  \sigma_\star^2 \rangle}.
\label{eqn:sigmaRMS}
\end{equation}

\subsection{Parametric Models}
If the physical state of the telescope can be described by a small
number of time-variable parameters $\{p_i\}$, then the PSF is some
function $PSF(x,y | p_i,x_\star,y_\star)$ of focal-plane position,
telescope state and the field position of the star.
We note that a great advantage of space-based observatories for WL work
is that the stability of their environment allows us to engineer a
telescope for which only a small number of degrees of freedom will
vary significantly.  For the SNAP telescope, the engineering
specifications are that all optical systems are stable to well below
the WL specification, except for:
\begin{itemize}
\item The alignment of the secondary mirror, which may vary due to
  imperfect performance of the feedback system that stabilizes the
  temperature of the mirror support structure;
\item The telescope line of sight (LOS) may vary during an exposure
  and smear the PSF due to noise and the finite bandwidth of the
  attitude control system (ACS) or due to high-frequency reaction wheel
  vibrations that transfer through the structure to optical elements,
  particularly the secondary mirror.
  % The pointing of the telescope, which may vary during an exposure
  % and smear the images, due to noise in the attitude control system
  % (ACS), or to high-frequency vibrations that transfer to optical
  % elements, particularly the secondary mirror.
\end{itemize}
In \S\ref{sec:PSF} we describe in detail the model we adopt for these
disturbances. 

Ground-based telescopes pose a more difficult challenge for PSF
modeling, because they have a very large number of time-varying
degrees of freedom.  Indeed the atmospheric distortions have infinite
degrees of freedom, formally.  Our analysis thus cannot be considered
valid for ground-based observatories.  \cite{Jarvis04} propose instead
that a principal-components analysis be performed on the ensemble of
PSF patterns observed by the telescope, so that the coefficients of
some finite number of principal components become the parameters for
the PSF model.
\cite{Jain06} discuss how changes in Seidel aberrations would be
manifested as PSF-change patterns, and might serve as a parameter set
for PSF modeling.  The success of these methods will depend upon how
well-behaved the telescope and atmosphere are, particularly whether
the optically significant perturbations are described by a small
number of variables.  An alternative approach, applied by
\cite{Wittman05} to PSF ellipticities induced by the atmosphere, is
to determine by {\it a priori} analysis that the disturbance will be
below the WL threshold.

%\subsection{Stellar ``Morphometry''}
\subsection{Simulations}

We simulate the following strategy:
\begin{enumerate}
\item Locate the stellar images in each exposure.
\item Measure PSF quantities at each star location; in our case, the
  second and third Gaussian-weighted moments.
\item Find the PSF parameters $\{\hat p_i\}$ that best reproduce the
  stellar data.
\item Use the $\{\hat p_i\}$ and the PSF model to derive the desired
  $\hat e_1$, etc., at any location in the focal plane.
\end{enumerate}
In the simulation we can then evaluate the RMS residual errors of the
PSF model.
%The use of PSF shapes to constrain telescope parameters has been
%christened ``morphometry.''

%In the simple case where the PSF has no spatial variation,
In the simple case where the PSF does not depend on the field position of stars,
we can
consider this procedure to be, essentially, averaging the measured
PSFs (and moments) of the observed stars.  In this case, we expected
the error on the PSF moments to be determined by the (quadrature) sum
of the signal-to-noise ($S/N$) levels of all the available stars.  In
particular, if all the stars are dominated by Poisson noise from
source photons, then $S/N=\sqrt N_\gamma$, where $N_\gamma$ is the
total number of photons collected from all usable stars in the
exposure.  We can therefore expect that
\be
e_1^{\rm RMS} \approx e_2^{\rm RMS} \approx \sqrt{2} \sigma_\star^{\rm RMS}
= \alpha N_\gamma^{-1/2}\,,
\label{eqn:sqrtn}
\ee
where $\alpha$ is some coefficient of order unity ($\alpha=1/\sqrt{2}$ for a
Gaussian PSF).

%More realistically, the PSF has spatial variation.
More realistically, the PSF does depend on the field position of stars. 
In this case, we
are using the PSF parametric model as a means of interpolating between
the stellar positions.  If the parameters are not too numerous, and
cause readily distinguished PSF patterns on the focal plane, then we
expect equation\,\ref{eqn:sqrtn} to continue to hold.  Our simulation will show
that this is indeed the case for the SNAP design, and we will aim to
estimate the coefficient $\alpha$.

\cite{PaulinHenriksson} instead consider the PSF to be locally
constant, and ask how large a region will contain enough stars to
adequately constrain this locally-constant PSF.  They then consider
this region size to be smallest on which WL observations can be
successful.  In reality both the time-varying PSF contamination and
the WL signal will have non-trivial angular power spectra and we have
to compare 
the bandpowers of each in setting our specifications.  \cite{Stabenau07}
investigate how PSF time variation will translate into multipole
patterns for a SNAP-like sky-scan strategy.  In this paper we will
simply calculate the RMS residual PSF errors, and note that they have
characteristic angular scales similar to the telescope field of view.

\section{Modeling the PSF}
\label{sec:PSF}

\subsection{Optical Aberrations and Diffraction}

  We adopt standard scalar diffraction theory to evaluate the optical
contribution to the PSF. The wavefront on the focal plane $\{x, y\}$
generated by a point source is,
\be
   U(x,y) = C \int \int d\xi d\eta P(\xi,\eta)
                 e^{ik \cdot OPD(\vec{\xi} , \vec{\theta})}
                 e^{-ik(\theta_x \xi + \theta_y \eta)}
                 e^{ik(x \xi + y \eta)/f} \,.
\label{eqn:waveFront}
\ee
Here $\vec{\xi}$ is the coordinate on the entrance pupil with components $\xi$
and $\eta$, $C$ is an uninteresting constant, $P(\xi,\eta)$ is the entrance
pupil function, $k = 2\pi/\lambda$ where $\lambda$ is the wavelength of
the band-limited optical light used to form the image, and $f$ is the
effective focal length of the telescope optics.
$OPD(\vec{\xi} ; \vec{\theta})$ is the optical path difference caused by the
lens/mirrors system
 which we expand using Zernike polynomials \citep{Noll76} as basis. 
The second exponential describes the phase differences
caused by the off-axis incident light ray in the direction $\vec{\theta}$ and
the third exponential is the phase differences caused by the
different distance light has to travel beyond the lens/mirrors and reach
the focal plane. The optical point spread function is
\be
   PSF(\vec{x}) = \left | U(x,y) \right |^2 \,.
\ee

\begin{figure}[ht]
\begin{center}
\includegraphics[width=3.6in]{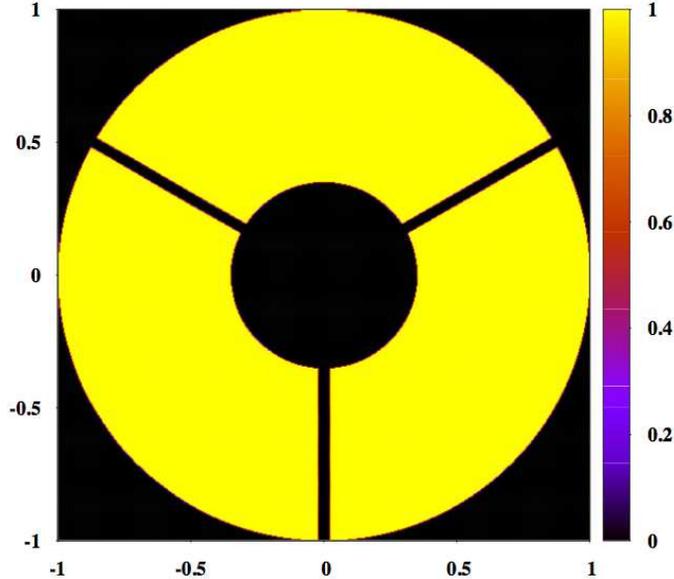}
\end{center}
\caption {Pupil function of SNAP telescope. The outer radius is 1 meter
          and the inner radius is 0.35 meter. The width of the three
          secondary mirror supporting struts is 4\,cm.
          }
\label{fig:pupil}
\end{figure}

     Figure\,\ref{fig:pupil} shows the pupil function of SNAP telescope.

\subsection{Optical Path Difference (OPD)}

The OPD map of the perfectly aligned telescope and the derivatives
with respect to misalignment parameters are calculated using ray
tracing through the telescope's optical system.  The OPD is projected
onto a Zernike basis, with results shown in Table\,\ref{tab:Jacobian}.
\be
  OPD(\vec{\xi},\vec{\theta}) = \sum_{n=1} \left [ C_{n}(\vec{\theta})
         + \sum_m D_m C^\prime_{mn}(\vec{\theta}) \right ] Z_n(\vec{\xi}) \,,
\label{eqn:opd}
\ee
where $Z_n$ is the $n^{th}$ Zernike polynomial, $D_m$ is the
the $m^{th}$ 
misalignment parameter, $C_{n}(\vec{\theta})$ is the OPD's
Zernike coefficients for pristine telescope and $C^\prime_{mn}(\vec{\theta})$ is
the Zernike coefficients of ${{\partial OPD} \over {\partial D_m}}$.

In the SNAP telescope design \citep{Lampton02,Sholl04}, the secondary mirror
position is expected to be the only optical dimension to vary significantly
with time due to thermal drift. The secondary mirror has 5 degrees of freedom
(DOF) which include shifts $D_x$ and $D_y$ in transverse directions, the
defocus $D_z$, plus tilts around the $x$ axis ($T_x$) and $y$ axis ($T_y$).

In practice we find that $C^\prime_{mn}(\vec{\theta})$ is a very weak
function of field for these parameters, and hence for the current
simulation we take $C^\prime_{mn}$ to be constants.  The pristine-telescope
Zernike coefficients $C_{n}(\vec{\theta})$ retain field dependence, but the
axisymmetry of telescope design reduces the freedom to the radial direction. 

\begin{table}[ht]
\caption{Zernike coefficients of the OPD in nm for a perfectly aligned
telescope with field location 10.4\,mrad off axis $C_{n}$, and the derivatives
$C^\prime_{mn}$ (see equation\,\ref{eqn:opd}).}
\centering
\begin{tabular*} {0.99\textwidth} [] {@{\extracolsep{\fill}} c c c c c c c c c c c}
\tableline\tableline
Zernike &  2 &  3  &  4  &   5  &   6  &   7   &   8   &   9  &  10  & 11  \\
\tableline
$C_{n}$ & 7.75 & 0.00 & 10.87 & 0.00 & 8.43 & 0.00 & -4.45 & 0.00 & 9.15 & -0.88 \\
\tableline
%Mirror  Partial derivative with respect to:
$D_x$ ($\mu {\rm m}$)     & -2.54 & 0.00 &  0.14 & 0.00 & 0.18 & 0.00  &  1.86  & 0.00 &  0.00 & 0.00 \\
$D_y$ ($\mu {\rm m}$)     & 0.00 & -2.55 &  0.00 & -0.18 & 0.00 & 1.86  &  0.00  & 0.00 &  0.00 & 0.00 \\
$D_z$ ($\mu {\rm m}$)     & -0.08 & 0.00 & -23.68 & 0.00 & 0.01 & 0.00  &  0.00  & 0.00 &  0.00 & 0.24 \\
$T_x$ ($\mu {\rm rad}$)  & 0.00 &  1.27 &  0.00 & 0.63 & 0.00 & -0.93  &  0.00  & 0.00 &  0.00 & 0.00 \\
$T_y$ ($\mu {\rm rad}$)  & -1.27 & 0.00 & -0.48 &  0.00 & 0.64 & 0.00 &   0.93 &  0.00&   0.00&  0.00 \\
\tableline
\end{tabular*}

%\tablenotetext{a}{Zernike coefficients for pristine telescope with field
%                  location of 10.4\,mrad off axis.}
%\tablenotetext{b}{$1\,\mu {\rm rad} \approx 0.2\,{\rm arcsec}$ }
%\tablecomments{OPD is in nm.}
\label{tab:Jacobian}
\end{table}

\subsection{Charge Diffusion}
The optical PSF must be convolved with the charge-diffusion pattern of the
CCD detector.
Charge diffusion is modeled as a Gaussian with fixed charge
diffusion length $\sigma_d=4\,\mu {\rm m}$. If the charge-diffusion
length were free to vary, it would be degenerate with an isotropic
telescope jitter (see below).
The PSF after charge
diffusion is
\be
  PSF(x,y) = \int dx' dy' PSF_0(x',y') {1 \over {\sqrt{2 \pi} \sigma_d}}
      {\rm exp} \left \{-{{(x-x')^2 + (y-y')^2} \over {2 \sigma_d^2}} \right \} \,,
\ee
where $PSF_0$ is the optical PSF. We execute the convolution with Fast Fourier
Transforms (FFTs).

\subsection{Jitter}

Guiding errors and mirror vibrations, also known as jitter, alter the effective 
PSF of a finite-length exposure. Each exposure hence has a unique
PSF map, even if the optics are otherwise stable. If the observatory is free
to rotate on all three axes, as for a space-borne observatory or 
an alt-az terrestrial telescope, then the effect of jitter varies across the
field of view, and is not a simple convolution of the image with a fixed
kernel. Stellar images in the exposure can
be used to infer the full field dependence of the jitter on the PSF.
We demonstrate here that as few as two stars are sufficient to fully reconstruct
the jittered PSFs, as long as the jitter amplitude is much less than the width
of the PSF. 
 
\subsubsection{Effect of Jitter on the PSF}
Assume that the modulation transfer function (MTF, the Fourier transform
of the PSF) at ${\bf x} = (x, y)$ from the optic axis is known to
be $T_0 ({\bf k})$ in the absence of telescope jitter. If the 
jitter has displaced the stellar image by some amount ${\bf \Delta x} = 
(\Delta x, \Delta y)$---which varies in time---then the transfer function
becomes
\be
  T({\bf k},t) = T_0({\bf k}) e^{-i {\bf k} \cdot {\bf \Delta x}(t)} \,.
\ee

The PSF for stellar images in the integrated exposure is the time-averaged
value
\be
  T({\bf k}) = {1 \over P} \int_0^P dt T({\bf k},t) = T_0({\bf k})
         \left < e^{-i{\bf k} \cdot {\bf \Delta x}(t)} \right > \,.
\ee

If ${\bf k} \cdot {\bf \Delta x} \ll 1$, 
then the exponential can be approximated by a Taylor expansion:
\be
  T({\bf k}) = T_0({\bf k}) \left [ 1 -
      i{\bf k} \cdot \left < {\bf \Delta x}(t) \right > 
   - {\bf k}^T \left < {\bf \Delta x \Delta x}^T \right > {\bf k}/2 \right ] \,.
\ee

The effect of the jitter on the PSF is then fully described by the mean
displacement $\left < {\bf \Delta x} \right >$ and by the covariance
matrix ${\bf C}_{\Delta x} = \left < {\bf \Delta x \Delta x}^T \right >$.
Further detail of the jitter history is irrelevant. The linear term is
simply a displacement of the entire PSF, and the quadratic term describes
a convolution of the jitter-free PSF with a very narrow jitter kernel. 
With a telescope of diameter $D$, focal length $f$, and wavelength $\lambda$,
% diffraction will force 
physical optics forces MTF=0 for 
$k < 2D/ (\lambda f)$. Therefore the Taylor expansion is valid if 
\be
  {{2D \Delta x} \over {\lambda f}} \ll 1 \,, 
\label{eqn:TaylorCond}
\ee
in other words the jitter must be much less than the size of the Airy
disk.

\subsubsection{Field Dependence of the Jitter MTF}

    If the observatory axis is displaced by angles 
$ \bft = (\theta_x , \theta_y , \theta_z ) $,
{\it{i.e.}} pitch, yaw, and roll, then 
the displacement of the image of star $i$ at $(x_i, y_i)$ is
\be
   \Delta x_i = f \theta_x - y_i \theta_z
\ee
\be
   \Delta y_i = f \theta_y + x_i \theta_z \,.
\ee

If the roll is non-zero, then the image displacement is field-dependent.
Now considering the observatory misalignment (jitter) to be a function of
time,
\be
  \left < \Delta x_i \Delta x_i \right > = f^2 \left < \theta_x \theta_x  \right >
         - 2 f y_i \left < \theta_x \theta_z  \right >
         + y_i^2 \left < \theta_z \theta_z  \right >
\label{eqn:Cxx}
\ee
\be
  \left < \Delta x_i \Delta y_i \right > = f^2 \left < \theta_x \theta_y  \right >
         + f x_i \left < \theta_x \theta_z  \right >
         - f y_i \left < \theta_y \theta_z  \right >
         - x_i y_i \left < \theta_z \theta_z  \right >
\label{eqn:Cxy}
\ee
\be
  \left < \Delta y_i \Delta y_i \right > = f^2 \left < \theta_y \theta_y  \right >
         + 2 f x_i \left < \theta_y \theta_z  \right >
         + x_i^2 \left < \theta_z \theta_z  \right >
\label{eqn:Cyy}
\ee

In the small-jitter limit, therefore, we find that the effect of the jitter
on the PSF at every point in the field of view is fully described by the six
independent elements of the jitter covariance matrix 
${\bf C}_{\theta} = \left < \bft {\bft}^T \right >$.
We can write
\be
  T ({\bf k}, {\bf x}) = T_0({\bf k}, {\bf x}) \left [ 1 -
        {\bf k} \cdot {\bf C}_{\Delta x}({\bf C}_{\theta},{\bf x})
          \cdot {\bf k} / 2 \right ] \,.
\ee

     In practice, therefore, if we have an exposure for which the
jitter-free PSF pattern is well determined, then we can completely determine
the jittered PSF anywhere in the focal plane by knowing 
the elements of ${\bf C}_{\theta}$. The jitter covariance matrix could be
determined from perfect knowledge of the PSF of any two stars in the
exposure. If there are a larger number of stars in the exposure, 
then the six jitter covariances are highly over determined, hence they are
easily derived for every exposure, even if there are other degrees of freedom
in $T_0$ which must be determined from these stars.

If the jitter amplitude is not small, then there can be a much larger
number of moments of the jitter history that are important, and a finite
number of stellar images may not in general recover full knowledge of the
effect of jitter on the PSF over the field. We will defer
consideration of this limit for another paper.

\section{Simulation Procedure}
\label{sec:procedure}

The simulation process is to: assume fiducial PSF parameters
(misalignment and/or jitter); create simulated stellar images across
the instrumented field of view; measure moments of these images; fit
the PSF model to these moments; and finally, evaluate the quality of the fit.

\subsection{Fiducial Model}
We analyze a fiducial case in which the secondary mirror is translated
1 $\mu$m and rotated 1 $\mu$rad from its correct position.  This would
be considered a very large error for the optomechanical system.  We
have verified that the choice of fiducial model does not influence the
RMS residuals to the fit.

When analyzing jitter, we assume an RMS motion of 36~mas in pitch and
yaw with 700~mas RMS in roll (as expected from SNAP telescope design). 
The RMS cross-correlation between axes is taken to be small by comparison.
Again, these do not strongly affect the results.

\subsection{Simulated Stars}
\begin{figure}[ht]
\centerline{ \psfig{file=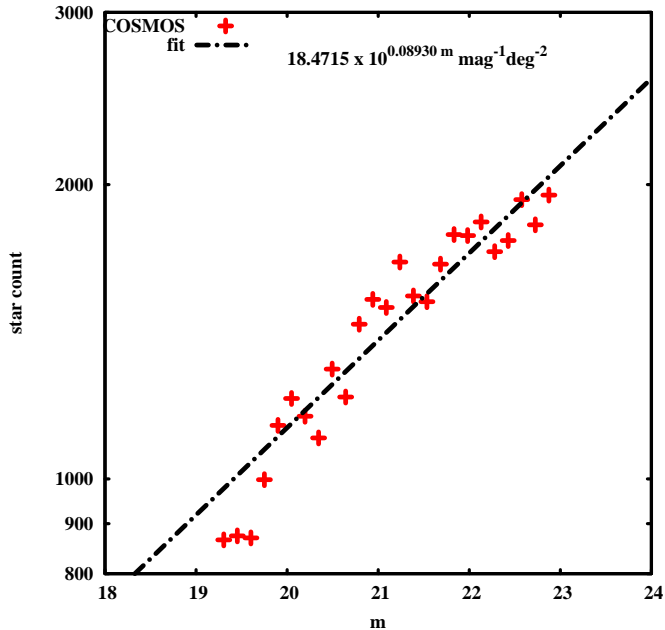, width=3.6in} }
\caption {COSMOS survey star magnitude distribution \citep{Robin06}.
          }
\label{fig:starDist}
\end{figure}

We use the star counts from the COSMOS HST survey as representative of
high galactic latitude fields.  
The star counts in the COSMOS field \citep{Robin06} are
well fit by (see Figure\,\ref{fig:starDist})
\be
   {dn^\star \over dm} = 18.5 \times 10^{0.089m} \,{\rm deg^{-2} mag^{-1}} \,,
\label{eqn:RobinStarDist}
\ee
where $m$ is stellar magnitude in the F814W band.

   If the star is too bright, it would saturate the SNAP CCDs.  We
hence conservatively assume that only stars with $19<m<23$ will be
used for morphometry.  The bright limit
roughly corresponds to 50,000 photons for a 300 second exposure.
The faint limit corresponds to $\approx 1200$ photons per star, which
is comparable to the sky background in the exposure.  Fainter stars
will contribute little to the PSF knowledge.

   The instrumented SNAP focal plane area is $\approx 0.7 \,{\rm deg^2}$.
This places approximately 2100 measurable stars on the focal plane, with
a total photon count of $N_\gamma\approx 2.3\times10^7$
per 300 sec exposure.  This suggests that an ideal morphometry
process would yield $e^{\rm RMS}$, $\sqrt{2} \sigma^{\rm RMS}_\star\sim
2\times10^{-4}$, well below the required weak lensing specification as
discussed in \S\,\ref{sec:intro}.

   We generate a sample of 2100 stars with random positions across the
focal plane. The magnitude of the stars is randomly generated according to
the magnitude distribution of equation\,\ref{eqn:RobinStarDist}.
The PSF, including optical distortions and charge diffusion is computed,
and is used to determine the mean number of photons detected in the CCD
pixels. The number of detected photons per pixel is drawn from a Poisson
distribution, and the quantum efficiency and gain of the SNAP pixels are
used to compute pixel values. Random dark noise of 5 photo-electrons
(as expected in the SNAP CCDs) is added to each pixel. The PSF moments
defined in equations\,\ref{eqn:m0}-\ref{eqn:pxy} and equation\,\ref{eqn:es}
are then computed from the pixel values. 

\subsection{Fitting Process}
For each available star in an exposure, we calculate the second and
third moments of a PSF image to which shot noise, sky noise, and read
noise have been added.  With the resulting
PSF moments as data, we search for the best fit misalignment parameters
by minimizing $\chi^2$
\begin{equation}
  \chi^2 = \sum_{i=1}^{N_\star} \sum_{jmom=1}^{Nmom}
           {{(M_i^{jmom} - \hat{M}_i^{jmom})^2} \over {\sigma_{moment}^2}}\,,
\end{equation}
where $N_\star$ is the total number of stars, $Nmom$ is the number of
independent PSF moments per star ($Nmom = 3$ if only 2nd moments are used),
$M$ is star moment (with noise), $\hat{M}$ is star moment calculated from
the model
(no noise), and $\sigma_{moment}$ is the rms of star moments. In general,
$\sigma_{moment}$ depends on star magnitude, filter band and the position
of the star on the focal plane. We assume a fixed filter band and
neglect the ($<10\%$) dependence on position. We produce a lookup
table of
$\sigma_{moment}$ vs source magnitude by Monte Carlo methods before
doing the fit.

\begin{figure}[ht]
\centerline{ \psfig{file=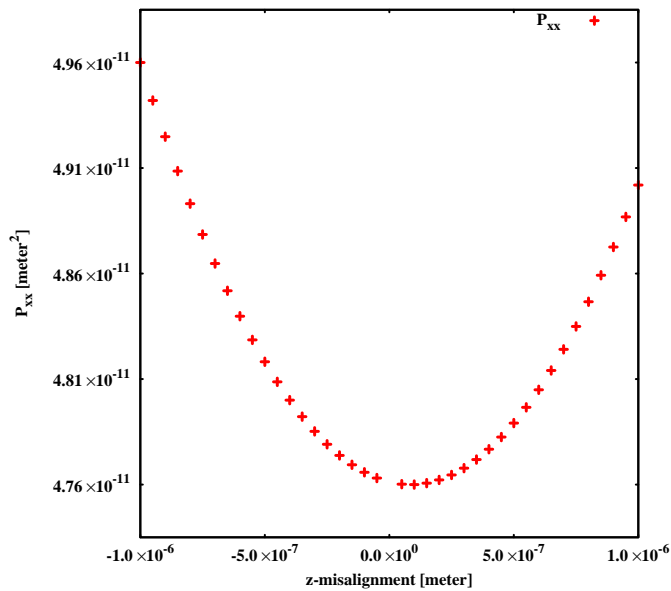, width=3.6in} }
\caption { Dependence of PSF moment $P_{xx}$ on $z$-misalignment (in meters).
           The star is located on the diagonal (i.e. $x=y$). }
\label{fig:testDer}
\end{figure}

The PSF moments depend nonlinearly on misalignment parameters. As an
example, Figure\,\ref{fig:testDer} shows the dependence of $P_{xx}$ on the
defocus parameter $D_z$.
The model-fitting procedure is hence nonlinear, so slower than a
linear $\chi^2$ fit.

\section{Results}
\label{sec:result}

     In this section, we show results of simulating and fitting for secondary
mirror misalignment only, and simulating and fitting with the jitter parameters
jointly.
We also show that the inclusion of third moments of the PSFs improves the fit.

\subsection{Fit for Secondary Misalignments Only}

\subsubsection{Using Second Moments of PSFs}

We first simulate a single exposure of SNAP, using only the PSF second
moments to constrain the secondary-mirror misalignment.
The fiducial misalignment parameters, their fitted values,
and the $1-\sigma$ uncertainties are listed in
Table\,\ref{tab:fit}. $D_z$ is precise 
to $<10$~nm which is well below the achievable
mechanical stability.  Hence the morphometric information can greatly
improve knowledge of defocus.
The other parameters are as much as 20 times less precise: accuracies of
$\approx0.2\,\mu$m in $D_x$ and $D_y$ seem quite disappointing, for
example. 

\begin{table}[ht]
\centering
\caption{Misalignment fitting results}
\begin{tabular*} {0.99\textwidth} [] {@{\extracolsep{\fill}} c c c c c c}
\hline\hline
misalignment & fiducial value & fitted value \tablenotemark{a} & $1-\sigma$ error \tablenotemark{b} &
fit incl. 3rd moments \tablenotemark{c} & $1-\sigma$ error \tablenotemark{d} \\
\hline
$D_x$ ($\mu {\rm m}$)             & 1   &  0.8542 &  0.1839 & 1.0052 & 0.1494 \\
$D_y$ ($\mu {\rm m}$)             &-1   & -1.3869 &  0.2020 &-1.0719 & 0.1602 \\
$D_z$ ($\mu {\rm m}$)             & 1   &  0.9870 &  0.0086 & 1.0005 & 0.0083 \\
$T_x$ ($\mu {\rm rad}$)  & 1   &  0.4500 &  0.3280 & 0.8104 & 0.3125 \\
$T_y$ ($\mu {\rm rad}$)  & 1   &  1.2370 &  0.3063 & 0.8549 & 0.2937 \\
\hline
\end{tabular*}

\tablenotetext{a}{Fit using second moments of the PSFs.}
\tablenotetext{b}{1-$\sigma$ error of the fit using second moments of the
                  PSFs.}
\tablenotetext{c}{Fit using both second and third moments of the PSFs.}
\tablenotetext{d}{1-$\sigma$ error of the fit using both second and third
                  moments of the PSFs.}
%\tablecomments{single band 885 nm.}

\label{tab:fit}
\end{table}

To better understand this wide range of precisions, we examine the
eigenvectors and eigenvalues of the parameter covariance matrix,
shown in Table\,\ref{tab:eigen}. The best-constrained eigenvector is along
the $D_z$ direction, {\it i.e.} defocus. Two eigenvectors are
$\approx 80$ times more poorly constrained, but this
is not hard to understand. If the secondary mirror were spherical with
radius $R$, it
would have only three degrees of freedom, as misalignments with
$D_x=-RT_y$ and $D_y=-RT_x$ leave the sphere invariant.  
A non-spherical secondary mirror breaks these degeneracies and
results in finite (but poor) constraints on these eigenvectors.

\begin{table}[ht]
\caption{Eigen values and eigen vectors of the misalignment covariance matrix}
\centering
\begin{tabular*} {0.99\textwidth} [] {@{\extracolsep{\fill}} c c c c c c}
\hline\hline
   & eigen 1 & eigen 2 & eigen 3 & eigen 4 & eigen 5 \\
\hline
$D_x$ ($\mu {\rm m}$)             & 
    0.0661 &  0.4852 &  0.7423 &  0.4573 &  0.0037 \\
$D_y$ ($\mu {\rm m}$)             &
   -0.5022 &  0.0535 &  0.4597 & -0.7304 & -0.0107 \\
$D_z$ ($\mu {\rm m}$)             &
    0.0033 &  0.0189 & -0.0058 & -0.0191 &  0.9996 \\
$T_x$ ($\mu {\rm rad}$)  &
   -0.8502 &  0.1516 & -0.2812 &  0.4185 &  0.0063 \\
$T_y$ ($\mu {\rm rad}$)  &
   -0.1438 & -0.8593 &  0.3981 &  0.2862 &  0.0245 \\
\hline
eigen values ($\lambda_i \times 10^{6}$)  &
    0.3781 &  0.3479 &  0.0797 &  0.0758 &  0.0052 \\
\hline
Residual ($\times 10^{6}$) \tablenotemark{a} &
    0.6180 & -0.3789 & -0.0370 &  0.0538 & -0.0071 \\
\hline
\end{tabular*}

\tablenotetext{a}{Projection of the vector that points from the fiducial
                  parameters to the fitted parameters onto the eigen vectors.}

\label{tab:eigen}
\end{table}

   The ``Residual'' row in the table gives  the projection of the
fitting error onto the eigenvectors. It shows that the fitted
parameters deviate from 
the fiducial mostly in the directions that have weak constraints. All the
deviations are at or below $1.6\,\sigma$ level which is a sign that the fitter
is working properly.

\begin{figure}[ht]
\centerline{ \psfig{file=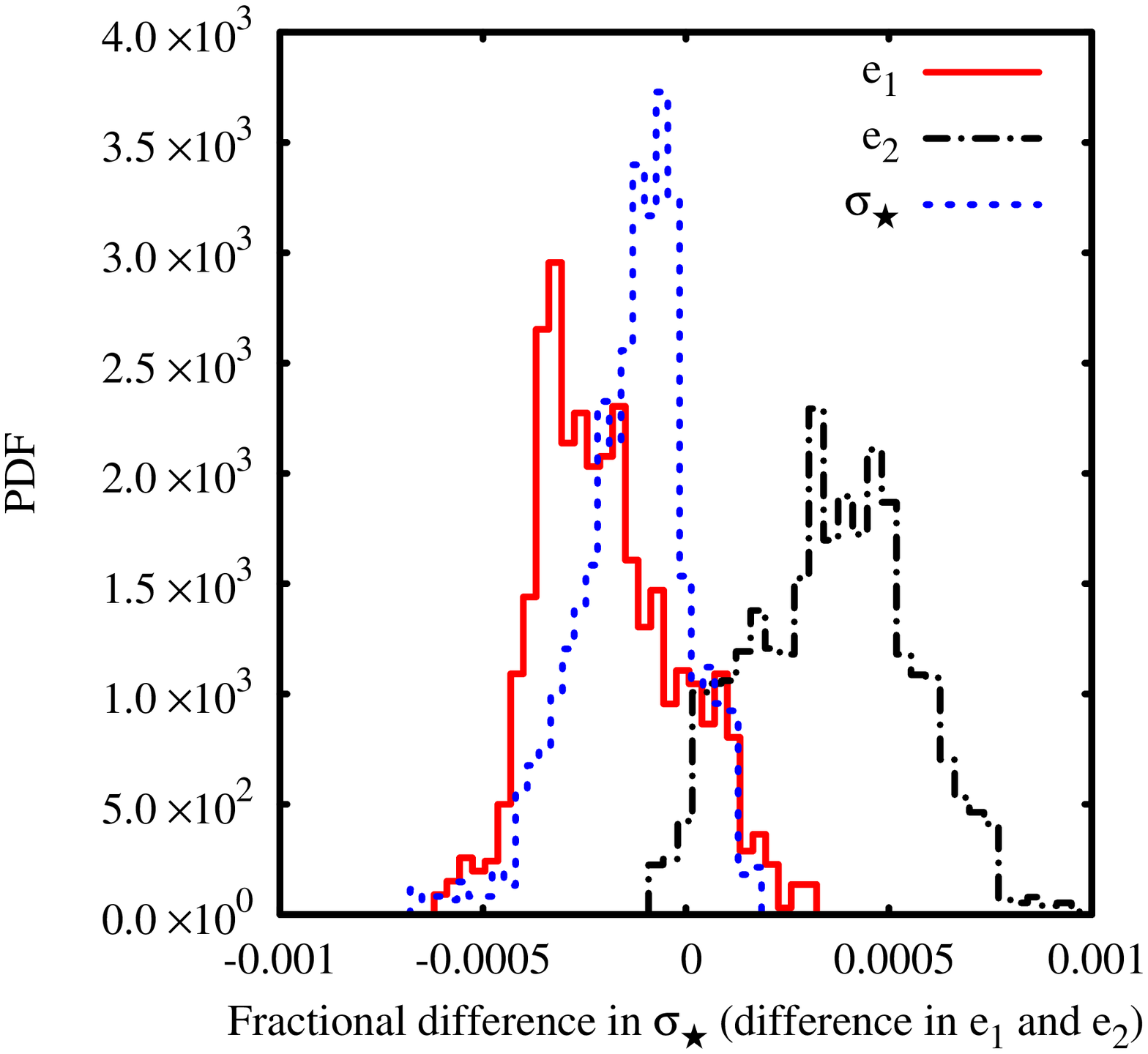, width=3.2in}
             \psfig{file=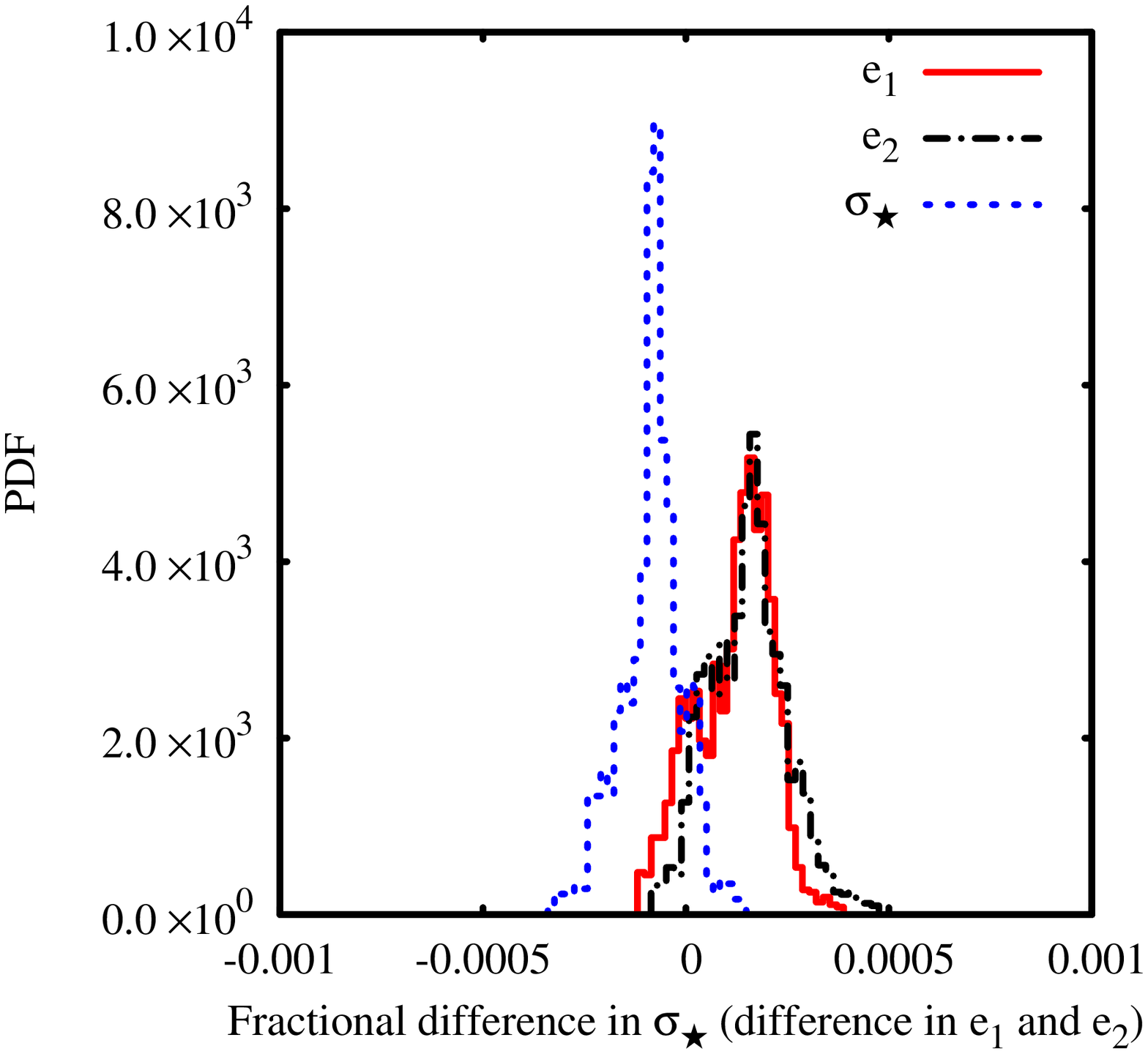, width=3.2in}
              }
\caption { Distributions of the residual PSF moments as defined in the text.
           Left panel: the fit is done using 2nd moments of the PSFs.
           Right panel: both 2nd and 3rd moments of the PSFs are used.}
\label{fig:momDiff_2ndvs3rd}
\end{figure}

The quantity that matters to weak lensing is the precision of the PSF knowledge.
The apparent poor fit to some of the misalignments reflects the insensitivity
of the PSF moments to certain combinations.  This insensitivity also
means, however, that poor knowledge of these eigenvectors has little
adverse effect on our PSF model. 
We compare the {\it noiseless} PSF ellipticities and size generated
using the fitted 
misalignment parameters with these generated using fiducial misalignment
parameters.  
Figure\,\ref{fig:momDiff_2ndvs3rd} (left panel) shows the distributions of the
residual PSF moments for a realization of a single exposure. We see
that the PSF errors are well below the $10^{-3}$ target for all three
quantities. The spread in PSF errors reflects the variation across the
focal plane. 

\subsubsection{Including Higher Moments of PSFs}

    There is information in the higher moments of the PSF. This could
further constrain the telescope misalignments. 
As shown in Table\,\ref{tab:fit}, the fitted misalignments are noticeably 
closer to the true values and the error bars are reduced when the
observed PSF second {\em and} third moments are used in the
misalignment fit.  Figure\,\ref{fig:momDiff_2ndvs3rd} shows the
reduction in PSF ellipticity and size residuals.
With the inclusion of third moments in the fit, the reduction of the
residual moments as shown in Figure\,\ref{fig:momDiff_2ndvs3rd} are much
more impressive than that of the one sigma errors shown in
Table\,\ref{tab:fit}. This is the manifestation that the contributions
to the one sigma errors are dominated by the two near degenerate eigens
which have little or no effect on the residual PSF moments.
In the following, we include third moments in the fit unless stated
otherwise.

\subsection{Fitting Misalignment and Jitter Parameters Jointly}

    Table\,\ref{tab:fitAll} shows the results of fitting secondary misalignment
and jitter parameters jointly. Adding the 6 jitter degrees of freedom to
the model roughly doubles the uncertainties on the
%three well-constrained components
5 misalignment degrees of freedom.
% (??? true???).
The parameter $D_z$ is still very precise (at $20\,{\rm nm}$ level).
The jitter parameters are determined without significant degeneracies
among each other nor with the misalignments; this is because these
parameters influence the PSF moments with rather distinct dependences
on field angle.
\begin{table}[ht]
\caption{Fitting for misalignment and jitter parameters}
\centering
\begin{tabular*} {0.99\textwidth} [] {@{\extracolsep{\fill}} l c c c}
\hline\hline
parameters & fit incl. 3rd moments & fiducial values & $1-\sigma$ error \\
\hline
$D_x$ ($\mu {\rm m}$)             & 0.5367 & 1   & 0.3596 \\
$D_y$ ($\mu {\rm m}$)             &-0.1932 &-1   & 0.3763 \\
$D_z$ ($\mu {\rm m}$)             & 0.9725 & 1   & 0.0203 \\
$T_x$ ($\mu {\rm rad}$)  & 2.6843 & 1   & 0.7503 \\
$T_y$ ($\mu {\rm rad}$)  & 1.9457 & 1   & 0.7207 \\
$\left < \theta_x \theta_x \right >$ ($\mu {\rm rad}^2$)
     &      0.0302      &       0.0300   &    0.0002 \\
$\left < \theta_x \theta_y \right >$ ($\mu {\rm rad}^2$)
     &      0.0013      &       0.0010   &    0.0001 \\
$\left < \theta_x \theta_z \right >$ ($\mu {\rm rad}^2$)
     &      0.0049      &       0.0010   &    0.0045 \\
$\left < \theta_y \theta_y \right >$ ($\mu {\rm rad}^2$)
     &      0.0298      &       0.0300   &    0.0002 \\
$\left < \theta_y \theta_z \right >$ ($\mu {\rm rad}^2$)
     &     -0.0011      &       0.0010   &    0.0046 \\
$\left < \theta_z \theta_z \right >$ ($\mu {\rm rad}^2$)
     &     19.5098      &      20.0000   &    0.9244 \\
\hline
\end{tabular*}
%\tablecomments{single band 885 nm; 2100 stars.}
\label{tab:fitAll}
\end{table}

\begin{figure}[ht]
\centerline{ \psfig{file=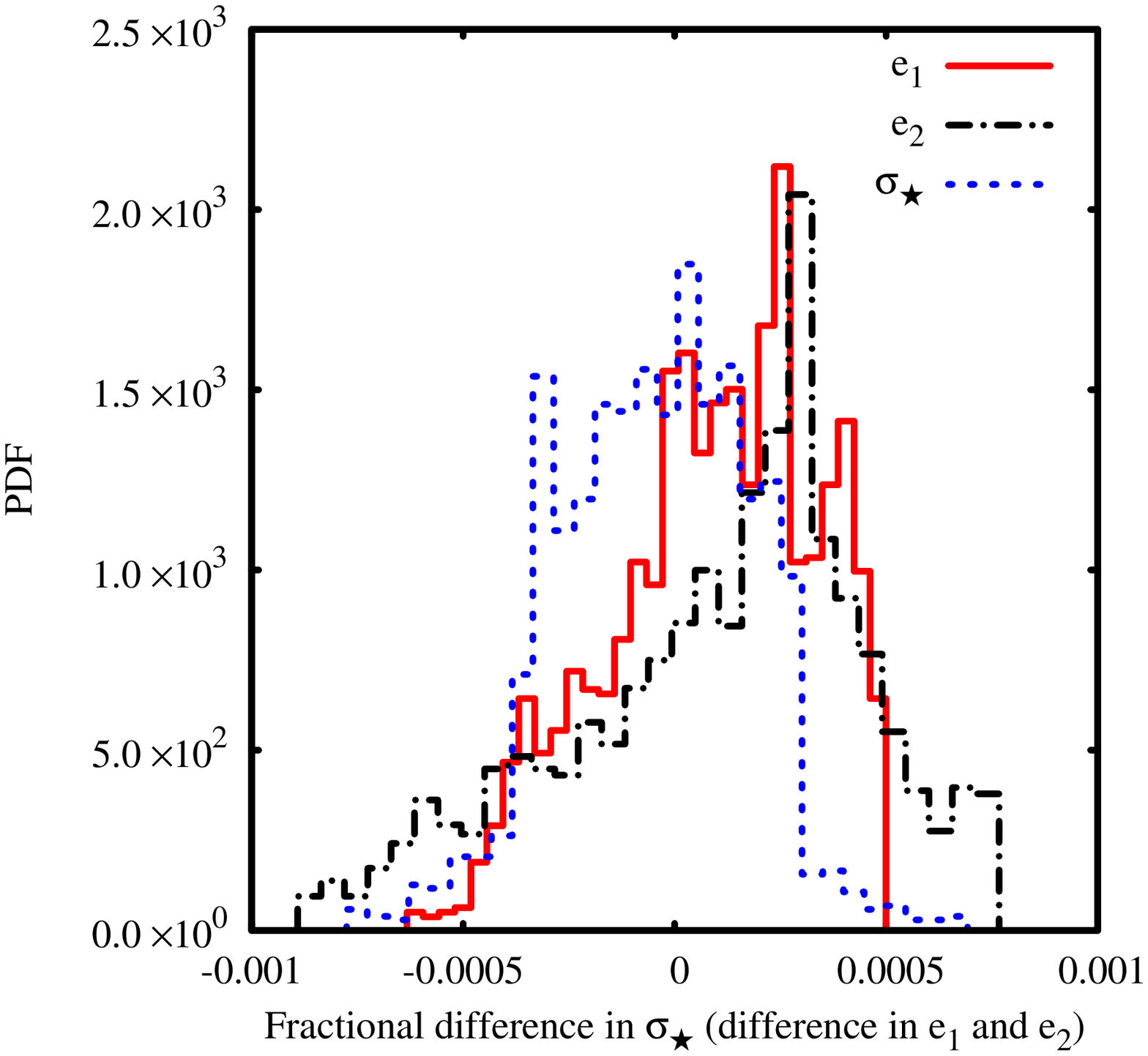, width=3.6in}
              }
\caption { Distributions of the residual PSF moments
           for fit including 5 secondary mirror misalignment and  6 jitter
           parameters. Both 2nd and 3rd moments of the PSF are utilized. }
\label{fig:diffMom_all}
\end{figure}

\begin{figure}[ht]
\centerline{ \psfig{file=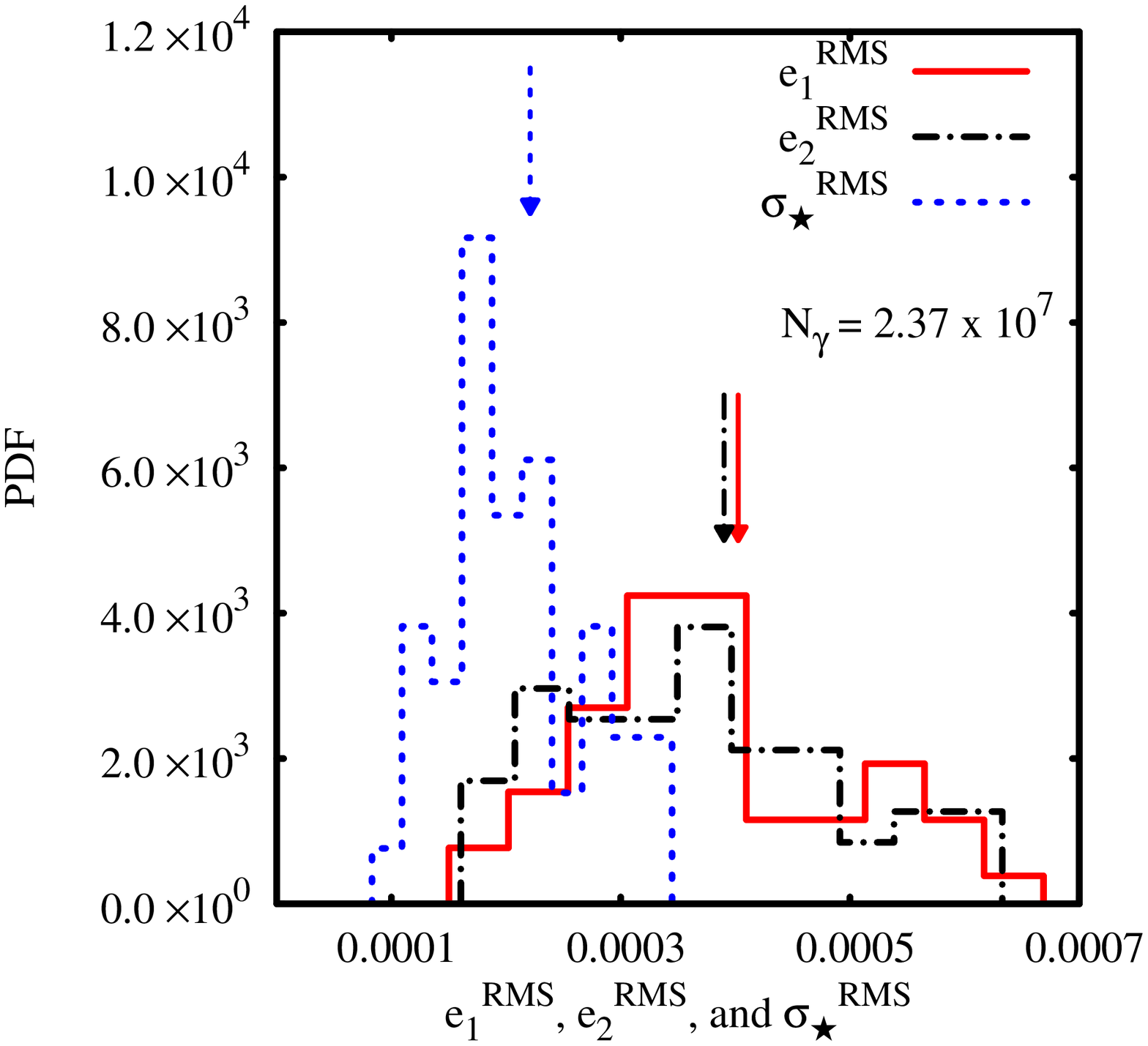, width=3.2in}
             \psfig{file=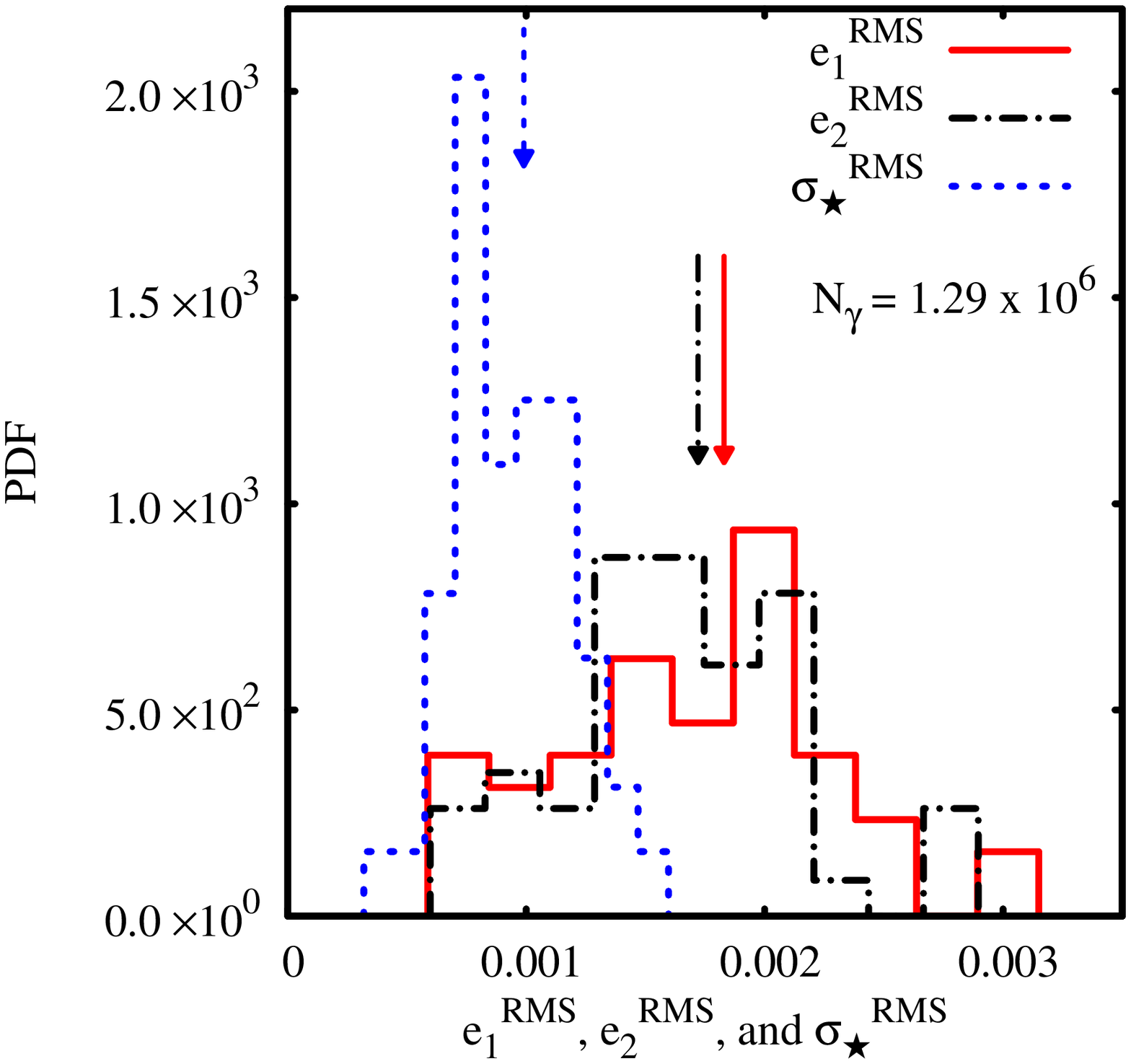, width=3.2in}
              }
\caption {Distributions of $e_1^{\rm RMS}$, $e_2^{\rm RMS}$, and
          $\sigma_{\star}^{\rm RMS}$ from 50 realizations.
          Left: 2100 stars are used for each realization.
          Right: 100 stars are used for each realization.
          The arrows point at the quadrature means of the distributions
          which are listed in Table\,\ref{tab:alpha}.
          Note that the horizontal scales are different in the
          two panels.
}
\label{fig:RMS}
\end{figure}

The residual PSF ellipticities and sizes for one realization of joint
misalignment/jitter fitting are shown in Figure\,\ref{fig:diffMom_all}.
As mentioned before, the spreads of the residual PSF $e_1, e_2$, and
$\sigma_\star$ distributions in this single realization are
due to field dependence. Different realizations of the data (star
locations and random noise change) produce different mean and spread of
the residual moments distribution.
Figure\,\ref{fig:RMS} left panel shows the distributions of $e^{\rm RMS}$
and $\sigma_\star^{\rm RMS}$ from 50 realizations.
We produce a single measure of the
efficacy of the morphometry procedure by averaging the RMS PSF
residuals over all focal-plane positions in many realizations, as per
equations\,\ref{eqn:e1RMS}-\ref{eqn:sigmaRMS}.
The mean values (in quadrature) of the distributions in
Figure\,\ref{fig:RMS} left panel
are exactly those. They are labeled by the arrows in the plot
and listed in Table\,\ref{tab:alpha} as well.
Calculated using equation\,\ref{eqn:sqrtn},
the $\alpha$ values are also tabulated in Table\,\ref{tab:alpha}.
Taking the average value, we find $\alpha \approx 1.8$. So we have
\be
e_1^{\rm RMS} \approx e_2^{\rm RMS} \approx \sqrt{2} \sigma_\star^{\rm RMS}
= 1.8 N_\gamma^{-1/2}\,.
\label{eqn:alpha}
\ee
%\begin{eqnarray}
%e_1^{\rm RMS} = e_2^{\rm RMS} & \approx & 3.0\times10^{-4} =  1.46 N_{\gamma}^{-1/2} \\
%\sigma_\star^{\rm RMS} & \approx & 1.8\times10^{-4} = 0.88 N_{\gamma}^{-1/2}.
%\end{eqnarray}
The ensemble
average residuals are consistent with zero, {\it i.e.} the PSF models
are unbiased, to the $2.5\times10^{-5}$ accuracy of our 50 realizations.

\begin{table}[ht]
\caption{Average $e^{\rm RMS}$ and $\sigma_{\star}^{\rm RMS}$ from 50
         realizations and $\alpha$ values}
\centering
\begin{tabular*} {0.99\textwidth} [] {@{\extracolsep{\fill}} l|c c|c c}
\hline\hline
   &  2100 stars per realization & $N_{\gamma} = 2.37 \times 10^7$ & 100 stars per realization & $N_{\gamma} = 1.29 \times 10^6$ \\
\hline
   & average & $\alpha$ & average & $\alpha$ \\
\hline
$e_1^{\rm RMS}$            & $4.0 \times 10^{-4}$ & 1.95 & $1.8 \times 10^{-3}$ & 2.04 \\
$e_2^{\rm RMS}$            & $3.9 \times 10^{-4}$ & 1.90 & $1.7 \times 10^{-3}$ & 1.93 \\
$\sigma_{\star}^{\rm RMS}$ & $2.2 \times 10^{-4}$ & 1.52 & $1.0 \times 10^{-3}$ & 1.14 \\
\hline
\end{tabular*}
%\tablenotetext{I}{Defined and calculated using equation\,\ref{eqn:sqrtn}.}
%\tablecomments{2100 stars: $N_{\gamma} = 2.37 \times 10^7$
%                           $N_{\gamma}^{-1/2} = 2.05 \times 10^{-4}$;
%               100 stars: $N_{\gamma} = 1.29 \times 10^6$
%                           $N_{\gamma}^{-1/2} = 8.80 \times 10^{-4}$}
\label{tab:alpha}
\end{table}

   To test our hypothesis that the PSF errors will scale as $N_\gamma^{-1/2}$, 
we repeat the simulation using 100 stars instead of the fiducial 2100
stars. The distributions of $e^{\rm RMS}$ and $\sigma_\star^{\rm RMS}$ are
shown in the right panel of Figure\,\ref{fig:RMS}. 
%The mean and std of the distributions and $N_{\gamma}$ are listed in the plot.
Again, the quadrature means of these distributions are labeled by the
arrows in the plot and listed in Table\,\ref{tab:alpha}. We find $\alpha
\approx 1.7$. So the RMS residuals do indeed scale as expected. 

From Figure~\ref{fig:diffMom_all} it is clear that part of the
residual errors in the PSF ellipticities are from a shift in the mean
across the field of view, while the rest is from errors that vary
across the field of view.  The two different types of PSF modeling
errors will propagate into different angular scales in the WL power
spectrum.  For the SNAP simulation, we find that roughly half of the
$e$ modeling variance is in the mean across the field of view.

Since essentially all the residual variance arises from shot noise in the
stars, it will be uncorrelated from exposure to exposure.

\section{Conclusion and Discussion}
\label{sec:conclude}

Our simulation of ``morphometry'' for the SNAP telescope demonstrates
that the $\approx2000$ well-measured stars in a typical exposure
contain sufficient information to reduce the errors in the modeled PSF
ellipticity and size to $4.0 \times10^{-4}$ and $2.2 \times10^{-4}$ 
respectively, giving significant margin to meet the $\approx10^{-3}$
level needed to reduce weak-lensing systematic errors below
statistical errors of future surveys.
For the SNAP telescope design and focal plane, we find
$e^{\rm RMS} \approx 1.8 / \sqrt{N_\gamma}$ 
and $\sigma_\star^{\rm RMS} \approx 1.1 / \sqrt{N_\gamma}$.

PSF estimation error in morphometry will be only part of the
shape-measurement error budget, so this margin is important.
Other potential source of errors in PSF estimation include charge
transfer inefficiency (CTI), data compression artifacts, and chromatic
PSF dependence that causes galaxies' PSFs to differ from stellar PSFs.
Shape measurement errors can also arise in the deconvolution process
even if the PSF is known precisely \citep{STEP1, STEP2}.
To have a successful weak lensing
mission, the sum of these errors must meet the weak lensing requirement.

We have also assumed that the only time-variable aspects of the PSF
are the secondary mirror alignment, and small pointing jitter.  The
SNAP spacecraft is engineered to take advantage of the
extremely stable space environment so that these are the only relevant
degrees of freedom.  
A ground based observatory would suffer in addition
the effects of wind, gravity loading, and seeing, which are
complicated to model, potentially involving a large number of
degrees of freedom.  We have seen in the SNAP case that adding 6
jitter degrees of freedom to the PSF model makes the PSF model errors
twice as large as when we fit for only 5 misalignment parameters,
so it seems likely that the PSF modeling performance will be degraded
by larger number of parameters in the system.

\acknowledgements {\it Acknowledgments}: We thank Michael Lampton,
Alexie Leauthaud and Mike Jarvis for useful discussions. ZM and GB are
supported by Department of Energy grant DOE-DE-FG02-95ER40893.
GB acknowledges additional support from NASA grant BEFS 04-0014-0018
and National Science Foundation grant AST 06-07667.
AW is supported by the Department of Energy under grant DE-FG02-92-ER40701,
and MS is supported by U.S. Department of Energy under contract
No. DE-AC02-05CH11231.

\end{document}